\documentclass[nohyperref]{article}

\usepackage{subfigure}
\usepackage[subfigure]{tocloft}

\usepackage{microtype}
\usepackage{graphicx}
\usepackage{subfigure}
\usepackage{booktabs} % for professional tables
\usepackage{makecell}

\usepackage{hyperref}
\newcommand{\indep}{\perp \!\!\! \perp}

\usepackage[accepted]{icml2022}

\usepackage{amsmath}
\usepackage{amssymb}
\usepackage{mathtools}
\usepackage{amsthm}
\usepackage[capitalize,noabbrev]{cleveref}

\theoremstyle{plain}

\theoremstyle{definition}

\theoremstyle{remark}

\usepackage[textsize=tiny]{todonotes}

\icmltitlerunning{COEM: Cross-Modal Embedding for MetaCell Identification}

\begin{document}

\twocolumn[
% \icmltitle{Submission and Formatting Instructions for \\
%           ICML Workshop on Computational Biology (ICML WCB 2022)}

\icmltitle{COEM: Cross-Modal Embedding for MetaCell Identification}


\icmlsetsymbol{equal}{*}

\begin{icmlauthorlist}
\icmlauthor{Haiyi Mao}{equal,yyy,comp}
\icmlauthor{Minxue Jia}{equal,yyy,comp}
\icmlauthor{Jason Xiaotian Dou}{sch}
\icmlauthor{Haotian Zhang}{yyy,comp}
\icmlauthor{Panayiotis V. Benos}{yyy,comp}
\end{icmlauthorlist}

\icmlaffiliation{yyy}{Department of Computational and Systems Biology, University of Pittsburgh, PA, USA}
\icmlaffiliation{comp}{Joint Carnegie Mellon University - University of Pittsburgh PhD Program in Computational Biology}
\icmlaffiliation{sch}{Department of Electrical and Computer Engineering, University of Pittsburgh, PA, USA}

\icmlcorrespondingauthor{Haiyi Mao, Minxue Jia}{[ham112, minxue.jia]@pitt.edu}

\icmlkeywords{Machine Learning, ICML}

\vskip 0.3in
]

\printAffiliationsAndNotice{\icmlEqualContribution} % otherwise use the standard text.

\begin{abstract}
Metacells are disjoint and homogeneous groups of single-cell profiles, representing discrete and highly granular cell states. Existing metacell algorithms tend to use only one modality to infer metacells, even though single-cell multi-omics datasets profile multiple molecular modalities within the same cell. Here, we present \textbf{C}ross-M\textbf{O}dal \textbf{E}mbedding for \textbf{M}etaCell Identification (COEM), which utilizes an embedded space leveraging the information of both scATAC-seq and scRNA-seq to perform aggregation, balancing the trade-off between fine resolution and sufficient  sequencing coverage. COEM outperforms the state-of-the-art method SEACells by efficiently identifying accurate and well-separated metacells across datasets with continuous and discrete cell types. Furthermore, COEM significantly improves peak-to-gene association analyses, and facilitates complex gene regulatory inference tasks. 
\end{abstract}

\section{Introduction}
\label{submission}
Recent advances in single-cell technologies have enabled measuring chromatin accessibility and gene expression levels simultaneously at single-cell resolution. Various sequencing technologies, such as sci-CAR~\cite{sci-CAR}, SNARE-seq~\cite{SNARE-seq}, SHARE-seq~\cite{share-seq}, and 10X Genomics Multiome, have been developed and applied to comprehensive studies of cell heterogeneity, developmental dynamics, and \textit{cis}-regulatory elements (CREs) \cite{chu2022cell}. However, extreme data sparsity is observed in Single-cell RNA sequencing (scRNA-seq) and Single-cell sequencing assay for transposase-accessible chromatin (scATAC-seq) data. This impedes \textit{cis}-regulation inference at the single-cell level. 

\begin{figure}[t!]
    \centering
    \includegraphics[scale=0.35]{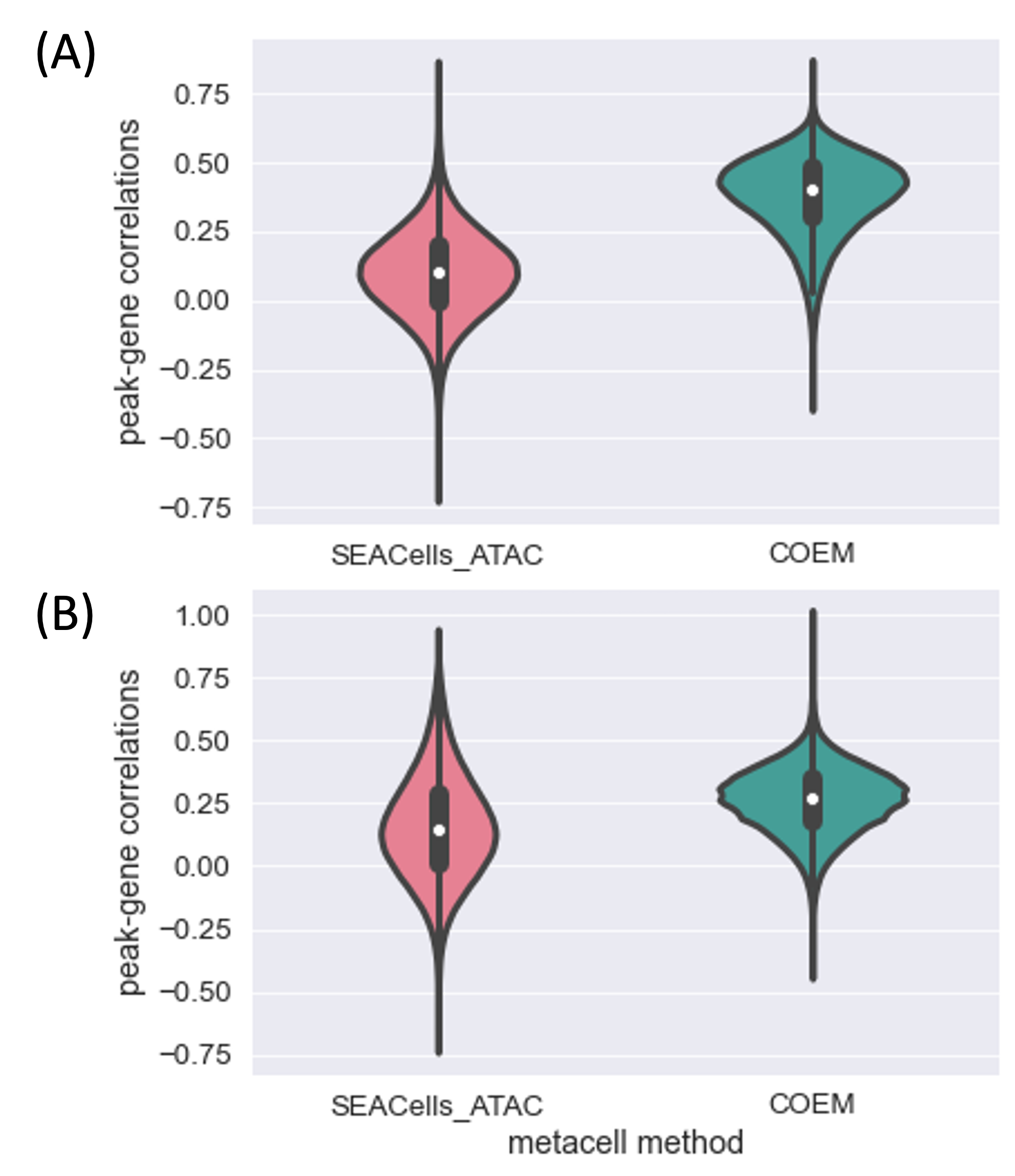}
    \caption{\small Violin plot of Pearson correlation between metacell-aggregated gene expression and accessibility on (A) CD34+ bone marrow dataset and (B) T-cell depleted bone marrow dataset. There are a considerable number of negative peak-to-gene associations in metacells identified by SEACell with scATAC-seq, while use of COEM-identified metacells improves peak-to-gene correlation.}
    \label{fig:cor}
\end{figure}
\vspace*{-3pt}

High-throughput single-cell profiling of biological samples typically leads to \textit{repetitive sampling} of highly similar and statistically equivalent cells. The concept of metacells \cite{baran2019metacell} has been proposed to maintain statistical confidence. Metacells are sets of scRNA-seq cell profiles representing distinct, highly granular cell states. Metacell aggregates, with their sufficient sequencing coverage, help reduce sparsity-derived problems in downstream analyses. 

The MetaCells algorithm~\cite{baran2019metacell} infers metacells based on the partition of  \textit{k}-nearest neighbor (KNN) similarity graphs on scRNA-seq data only, but it fails on scATAC-seq data. The SEACells approach (Single-cell aggregation of cell-states) \cite{Persad2022.04.02.486748} is designed to identify metacells in RNA or ATAC modality. It significantly improves peak-to-gene associations, i.e. between chromatin accessibility peak and gene expression. However, SEACells identifies metacells based on only one modality: holding onto a strict biological assumption that accessible chromatin is associated with active transcription consistently, despite different data modalities possibly exhibiting a time lag between chromatin remodeling of a gene and its transcription. The occurrence of this "time lag" manifests when: 
 \begin{enumerate}
     \item Chromatin opens before transcription initiates
     \item mRNA degrades follows chromatin closing ~\cite{MultiVelo}.
 \end{enumerate}
 So, it is not surprising that we observe negative peak-to-gene associations across metacells, as Fig.~\ref{fig:cor} shows.  We aim to investigate that  integrated analysis of scATAC-seq and scRNA-seq data to improve our ability to discover and characterize cell states, while avoiding biased associations.

Our new algorithm, COEM, is designed to integratively analyze multi-omic sequencing single-cell data (e.g., gene expression and chromatin accessibility measured in the same cell). COEM addresses the fusion challenge \cite{baltruvsaitis2018multimodal} by first learning joint low-dimensional latent representations for scRNA-seq and scATAC-seq data through a multi-view, multi-modal variational auto-encoder (VAE) model; further utilizing a spectral clustering approach, based on graph connectivity, to identify metacells representing highly granular, distinct cell states on \textit{both} RNA and chromatin levels. We found COEM to be
% \textcolor{red}{[significantly?]} 
more accurate than SEACells in metacell identification across datasets with discrete 
%\textcolor{red}{[discrete or distinct?]}
cell types and continuous developmental trajectories. Also, COEM-identified metacells substantially improve the peak-to-gene association analysis and empower CREs prediction. 
\begin{figure*}[t!]
    \centering
    \includegraphics[scale=0.45]{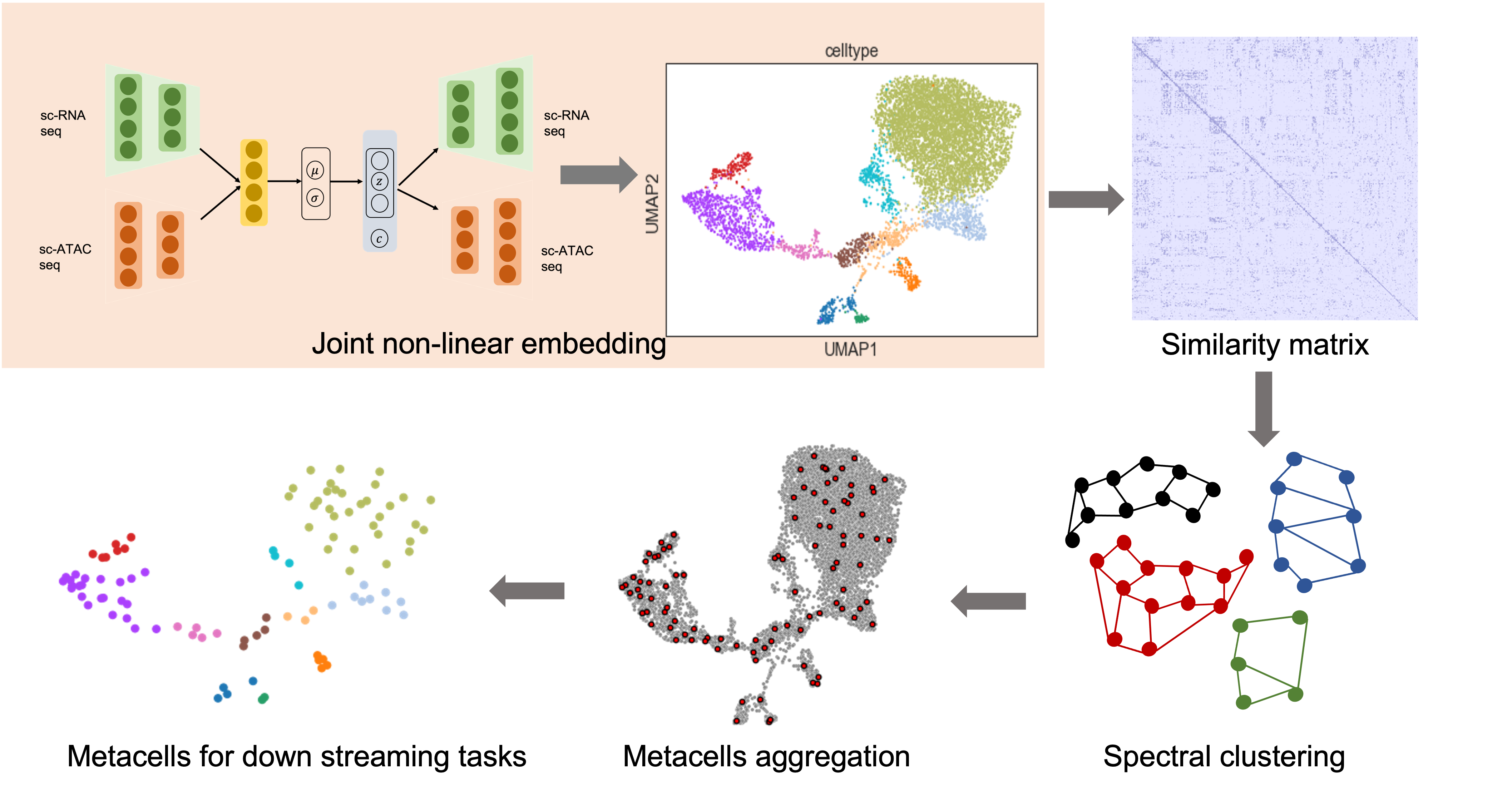}
    \caption{\small The main steps of COEM. 1. The joint profiled scRNA-seq and scATAC-seq are fed to the modified VAE to extract a common latent representations $\mathbf{z}$. 2. Then a similarity matrix are constructed based on $\mathbf{z}$ and KNN graphs. 3. Spectral clustering is applied to identify the higher granularity clusters where Metacells belongs to. 4. The Metacells features are aggregated for each clusters. 5. Metacells are availabe for downstreaming tasks. }
    \label{fig:pipelines}
\end{figure*}
\vspace*{-3pt}
\section{Method}
\subsection{Multi-Modal Joint Embedding}
COEM learns a low dimensional representation from the joint scRNA-seq and scATAC-seq data, measured in the same cell. We adopt a modified multi-view VAE to extract a common latent space of the two data modalities. This is similar to previous work in VAEs representing the single-cell sequence data \cite{lopez2018deep, li2022deep,ding2018interpretable,mao2020interpretable,xu2020learning} in less noisy low dimensions. The generative model of the VAEs is derived as follows. Given a cell, the scRNA gene expression $\mathbf{x} \in  \mathcal{P}_{\mathbf{x}}$, scATAC chromatin accessibility data $\mathbf{y} \in \mathcal{P}_{\mathbf{y}}$, cell type $c$, and the common latent representation $\mathbf{z}$, we have the following probability densities,
\begin{equation}
    \begin{aligned}
    \setlength{\belowdisplayskip}{1pt}
    & p_{\theta}(\mathbf{x}, \mathbf{z}| c) = p_{\theta}(\mathbf{x}|\mathbf{z}, \mathbf{c})p_{\theta}(\mathbf{z}|c)]\\
    & p_{\theta}(\mathbf{y}, \mathbf{z}|c) = p_{\theta}(\mathbf{y}|\mathbf{z}, \mathbf{c})p_{\theta}(\mathbf{z}|c),
    \end{aligned}
\end{equation}
where $\theta$ represents the parameters of the generative models. The likelihood $p_{\theta}$ of $\mathbf{x}$ and $\mathbf{y}$ depends on the common latent space $\mathbf{z}$ and the cell type $c$. With the assumption that the distribution of  observed scRNA-seq $\mathbf{x}$, scATAC-seq $\mathbf{y}$ are independent given the common latent representations $\mathbf{z}$ and cell type $c$, $\mathbf{x} \indep \mathbf{y} | \mathbf{z}, c$. So, we have following joint generative model,
% \begin{small}
\begin{equation}
    \begin{aligned}
    p_{\theta}(\mathbf{x}, \mathbf{y}, \mathbf{z}| c) &= p_{\theta}(\mathbf{x}, \mathbf{y}|\mathbf{z}, c)p_{\theta}(\mathbf{z}|c)\\
    & = p_{\theta}(\mathbf{x}, |\mathbf{z}, c)p_{\theta}(\mathbf{y}, |\mathbf{z}, c)p_{\theta}(\mathbf{z}|c).
    \end{aligned}
\end{equation}
% \end{small}
Moreover, we have the following Evidence Lower Bound (ELBO) for two modalities with conditional variational approximation $q_{\phi}$, 
% \begin{small}
\begin{equation}
    \begin{aligned}
     \log p_{\theta}(\mathbf{x}|c) \geq &-KL(q_{\phi}(\mathbf{z}|\mathbf{x}, c)) || p_{\theta}(\mathbf{z} | c)) \\
    & + \mathbb{E}_{q_{\phi}(\mathbf{z}|\mathbf{x}, c))}(\log p_{\theta}(\mathbf{x}|\mathbf{z}, c)) \\
%     \end{aligned}
% \end{equation}   
% \begin{equation}
%     \begin{aligned}
    \log p_{\theta}(\mathbf{y}|c) \geq
    & -KL(q_{\phi}(\mathbf{z}|\mathbf{y}, c)) || p_\phi(\mathbf{z} | c))\\
    & + \mathbb{E}_{q_{\phi}(\mathbf{z}|\mathbf{y}, c)}(\log p_{\theta}(\mathbf{y}|\mathbf{z}, c)).
    \end{aligned}
\end{equation} 
Normally, we use an inference network (encoder)  to learn the variational approximation $q_{\phi}$ and a generative network (decoder) to learn $p_{\theta}$. Particularly, we assume $p_\theta(\mathbf{x}|\mathbf{z}, c)$ is a zero-inflated Poisson (ZIP) distribution, $p_{\theta}(\mathbf{y}|\mathbf{z}, c)$ is a negative binomial (NB) distribution, and $p_\theta(\mathbf{z}|c)$ is a Gaussian Mixture Model $\mathbf{z} \sim \mathcal{N}(\mathbf{u}_{c}, \Sigma_{c})$~\cite{li2022deep,lopez2018deep}.
Given the proposed joint-embedding generative model, we adopt the scMVP \cite{li2022deep} architecture to model the generative
network and inference network. Specifically, there is a two-channel attention-based encoder network for the scRNA-seq and scATAC-seq inputs, then concatenating them to derive the posterior distribution $p_{\phi}(\mathbf{z}|\mathbf{x}, \mathbf{y}, c)$. 
% parameters of common latent embedding $\mathbf{z}$ with a Gaussian mixture model prior $p_{\theta}(\mathbf{z}|c)$.
Next, the imputed scRNA and scATAC profiles are reconstructed by an attention based two-channel decoder network to compute the likelihood $p_{\theta}(\mathbf{x}, \mathbf{y} | \mathbf{z}, c)$.
\subsection{Metacells Identification}
Similar to previous work \cite{baran2019metacell, Persad2022.04.02.486748}, COEM is a graph-based approach that employs spectral clustering to compute metacells. To be more specific, COEM consists of the following four steps:
\vspace{-1mm}
\begin{enumerate}
    \vspace{-0.2cm}\item Apply cross-modal joint embedding VAE  to learn low common representations $\mathbf{z}$ to encode the common information of scRNA-seq and scATAC-seq. 
    \vspace{-0.2cm}\item Build a \textit{k}-nearest neighbors (KNN) graph on the low dimensional space $\mathbf{z}$ based on Euclidean distances. 
    \vspace{-0.2cm}\item Then similarity matrix is constructed from the previous KNN graph by connectivity. The similarity matrix is based on radial basis function (RBF) kernel to encode the non-linear relations in the $\mathbf{z}$.
    \vspace{-0.2cm}\item Utilize spectral clustering to identify the cluster and to subsequently aggregate metacells as in  \cite{Persad2022.04.02.486748}.
\end{enumerate}
\vspace{-2mm}
This is one of the differences between COEM and SEACells~\cite{Persad2022.04.02.486748}, which uses archetypal analysis for metacell identification, instead of spectral clustering. While they both are graph based methods, instead of the convexity assumption of archetypal analysis of SEACells, spectral clustering of COEM purely utilizes the graphs' connectivity information and enjoys a faster running speed. We empirically validate that spectral clustering can improve metacell identification and running time.
\begin{small}
% \vspace{-1 mm}
\begin{table}[t!]
\setlength\tabcolsep{2pt}
\vskip 0.15in
\begin{center}
\begin{small}
\begin{sc}
\begin{tabular}{lccccr}
\toprule
methods & CD34 & sciCAR & 10xPBMC & Snare cellline\\
\midrule
Spectral &\textbf{23.7}&\textbf{6.3}& \textbf{45.2} &\textbf{1.7}\\
Archetypal &191.5&103.6&277.4 & 28.0\\
\bottomrule
\end{tabular}
\end{sc}
\end{small}
\end{center}
\vskip -0.1in
\caption{Running Time (seconds). We use the same low dimensional embeddings $\mathbf{z}$ to compare archetypal analysis (used in SEACells) and spectral clustering (used in COEM).}
\label{tab:time}
\vspace{-5.12mm}
\end{table}
\end{small}
\begin{small}
\begin{table}[h]
\begin{center}
\begin{sc}
\begin{tabular}{lcccr}
\toprule
   &\textbf{RNA\_pca}&\\
methods & separation & compactness\\
\midrule
% COEM+linear& 0.115 & \textbf{0.045}\\
% COEM+rbf &0.096&0.053\\
COEM & \textbf{0.235 $\begin{tiny}\pm 0.015\end{tiny}$} & 0.193 $\begin{tiny}\pm 0.017\end{tiny}$\\
SeaC-ATAC    & 0.196 $ \begin{tiny}\pm 0.007\end{tiny}$ & 0.686 $\begin{tiny}\pm 0.035\end{tiny}$ \\
SeaC-RNA    &  0.177 $\begin{tiny}\pm 0.051\end{tiny}$
 & \textbf{0.150 $\begin{tiny}\pm 0.062\end{tiny}$}\\
\toprule
&\textbf{ATAC\_svd}&\\
methods & separation & compactness\\
\midrule
% COEM+linear& 0.048 & 0.019 \\
% COEM+rbf & 0.040 & 0.018 \\
COEM & \textbf{0.091 $\begin{tiny}\pm 0.006\end{tiny}$} & \textbf{0.028 $\begin{tiny}\pm 0.002\end{tiny}$} \\
SeaC-ATAC    &  0.086 $\begin{tiny}\pm 0.002\end{tiny}$ & 0.004 $\begin{tiny}\pm 0.001\end{tiny}$ \\
SeaC-RNA    &  0.041 $\begin{tiny}\pm 0.015\end{tiny}$ & 0.032 $\begin{tiny}\pm 0.014\end{tiny}$\\
\bottomrule
 methods & Cell-type Purity \\
\midrule
% COEM+linear& 0.685 \\
% COEM+rbf & 0.669 \\
COEM &\textbf{0.907 $\begin{tiny}\pm 0.006\end{tiny}$}  \\
SeaC-ATAC    &  0.666 $\begin{tiny}\pm 0.01\end{tiny}$ \\
SeaC-RNA    &  0.734 $\begin{tiny}\pm 0.002\end{tiny}$ \\
\bottomrule
\end{tabular}
\end{sc}
\end{center}
% \vspace{-5 mm}
\caption{Metacacell Evaluation on sci-CAR celline dataset. Greater separation, lower compactness, and higher cell-type purity indicate better performance. COEM identified metacells with higher purity and separation}
\label{tab:sciCAR}
\end{table}
\end{small}
\vspace{-3 mm}

\begin{small}
\begin{table}[h]
\vskip 0.15in
\begin{center}
\begin{small}
\begin{sc}
\begin{tabular}{lcccr}
\toprule
&\textbf{RNA\_pca}&\\
methods & separation & compactness\\
\midrule
% COEM+linear&0.674&0.547\\
% COEM+rbf &0.776&0.799\\
COEM &\textbf{0.801 $\begin{tiny}\pm 0.022\end{tiny}$}&0.419 $\begin{tiny}\pm 0.001\end{tiny}$\\
SeaC-ATAC &0.723 $\begin{tiny}\pm 0.024\end{tiny}$ &1.061 $\begin{tiny}\pm 0.140\end{tiny}$\\
SeaC-RNA &0.704 $\begin{tiny}\pm 0.041\end{tiny}$ &\textbf{0.292 $\begin{tiny}\pm 0.217\end{tiny}$}\\
\toprule
&\textbf{ATAC\_svd}&\\
methods & separation & compactness\\
\midrule
% COEM+linear &0.560&0.577\\
% COEM+ rbf &0.601&0.817\\
COEM &\textbf{0.710 $\begin{tiny}\pm 0.013\end{tiny}$}&0.662 $\begin{tiny}\pm 0.019\end{tiny}$\\
SeaC-ATAC   &0.665 $\begin{tiny}\pm 0.056\end{tiny}$ &0.640 $\begin{tiny}\pm 0.153\end{tiny}$\\
SeaC-RNA    &0.566 $\begin{tiny}\pm 0.070\end{tiny}$ &\textbf{0.547 $\begin{tiny}\pm 0.291\end{tiny}$}\\
\bottomrule
\end{tabular}
\end{sc}
\end{small}
\end{center}
% \vspace{-1mm}
\caption{Metacacell Evaluation on T-cell depleted bone marrow dataset. COEM metacells have higher separation than SEACell's}
\label{tab:BM}
\end{table}
\end{small}

\section{Experiments}
\subsection{Evaluation Metrics}
As in the SEACells paper \cite{Persad2022.04.02.486748}, we use compactness, separation, and cell-type purity for the metacell benchmark. Compactness measures how homogeneous the cells within a metacell are. Separation assesses how metacells are distinct from each other. Cell-type purity, instead, measures the consistency of cell-types among cells that constitute a metacell.  
We computed diffusion components using principal components from RNA assay (RNA\_PCA) or singular values singular values from ATAC assay (ATAC\_SVD), and then quantified the compactness and separation of metacells.
Since SEACells identifies metacells from one modality, we denote them as SeaC-ATAC and SeaC-RNA.

\subsection{Results}
% \textcolor{red}{[Tables 2 and 3 are not descriptive. The legend should clearly say what we see, not just the dataset name. What are the ATAC and RNA lines? Which method do you compare to? SEACells? It should be noted like SeaC-ATAC, SeaC-RNA and the notation explained in the legend.]}

We use five datasets for evaluation: sci-CAR cell line~\cite{sci-CAR}, SNARE-seq cell line~\cite{SNARE-seq}, 10X Genomics Multiome PBMC, 10X Multiome CD34+ bone marrow (hematopoietic stem and progenitor cells) and 10X Multiome  T-cell depleted bone marrow~\cite{Persad2022.04.02.486748}. As showed in Table~\ref{tab:sciCAR} and \ref{tab:BM}, and Table \ref{tab:SNARE} ,\ref{tab:CD34}, and \ref{tab:10XPBMC} in appendix, COEM outperforms SEACells in most cases.\par
% \textcolor{red}{[can we calculate any p-value for those results?]}
The sci-CAR dataset profiles well-labeled cells, including cell lines 293T, 3T3, a 293T/3T3 cell mixture, and A549 cells after 0, 1, or 3 hours of dexamethasone treatment. We evaluate cell-type purity in sci-CAR dataset, and metacells from COEM algorithm has greater purity (0.91)
% \textcolor{red}{["greater purity" is vagus. You need to put numbers here.]} 
than SeaC-ATAC (0.67) and SeaC-RNA (0.74), which suggests that leveraging two modalities improves our ability to define cellular states, whereas achetypal analysis on joint embedding representations cannot guarantee high purity of metacells shown in Table~\ref{tab:sciCAR}.

Then, we measure compactness and separation across datasets with discrete cell types. Generally, COEM-based methods separate metacells better than SeaC-ATAC and SeaC-RNA shown in \cref{tab:BM,tab:sciCAR,tab:SNARE,tab:CD34,tab:10XPBMC}. 
%textcolor{red}{[need specific numbers here to demonstrate this.]}
But they may have higher compactness based on only one modality, since COEM balances cell variability from two modalities at the same time. The metacells identified by SEACells have lower compactness because metacell identification and compactness evaluation are based on the diffusion components from a single modality. But those metacells may not be well-separated and compacted based on the diffusion components from the other modality. Therefore, the association analysis between gene expression and peak accessibility may be biased using metacells inferred from only one modality. Further, spectral clustering outperforms archetypal analysis for identifying metacells with high cell-type purity and separation based on low dimensional embedding $\mathbf{z}$. We also measure running time with the same low dimensional embedding $\mathbf{z}$ for archetypal analysis and spectral clustering shown in Table~\ref{tab:time}. We can see spectral clustering has generally one order of magnitude shorter running time in the four datasets.

Analyzing chromatin accessibility and gene expression jointly allows us to infer the relationship between open-chromatin peaks and active transcription, aiding COEM in CRE discovery. Metacells can be used as a wise strategy to have high resolution and sufficient sequencing coverage for gene regulation inference from single-cell data. We evaluate the co-variation in chromatin accessibility and gene expression across metacells by computing Pearson correlations for each peak within +/- 100kb of the gene. We observe a large portion of peaks is negatively correlated with gene expression in SeaC-ATATC metacells (CD34+ bone marrow: 25.1\%; T-cell depleted bone marrow: 23.0\%) (Figure \ref{fig:cor}), while a small portion of negative peak-to-gene associations is identified in COEM metacells (CD34+ bone marrow: 1.3\%; T-cell depleted bone marrow: 3.0\%), which is consistent with reported ratio of negative peak-to-gene associations (1.2 $\sim$ 11 \%) at the single-cell level. This shows that the bias from the “time lag” phenomenon during differentiation is exaggerated in SEACells metacells but not in COEM. Besides, we could observe stronger peak-to-gene correlations for core genes based on COEM metacells. 
For example, GATA2 gene acts as a master regulator of proliferation and maintenance of hematopoietic stem and progenitor cells. The peak-to-gene correlation of GATA2 in COEM metacells from CD34+ bone marrow dataset is 0.73, while its correlation is 0.62 in SeaC-ATAC metacells. (The peak-to gene correlation of GATA2 is 0.1 at single-cell level) % both correlations are significant 
% For example, hematopoietic differentiation involves upregulation of erythroid gene KLF1 and downregulation of stem gene LPCAT2. The peak-to-gene correlations of KLF1 and LPCAT2 in COEM metacells on CD34+ bone marrow dataset are 0.74 and 0.70, while their correlations are 0.65 and 0.61 in SEACells metacells. \textcolor{red}{[significance?]}

% Lastly, it is worth noting that linear kernel with achetypal analysis performs relatively similar to RBF kernel, mainly because the cross-modal embedding representations already capture the non-linear relations amongst cells.

\section{Discussion}
Our experiments show that integrating information from both scRNA-seq and scATAC-seq is beneficial for identifying accurate and robust metacells from single-cell data. Especially joint-embedding, low-dimensional representations can enable the identification of cellular states with high cell-type purity and well-separated metacells. On top of that, the proposed COEM algorithm enjoys a faster running time. Moreover, it avoids high numbers of negative peak-to-gene links that are false.  Hence, the joint profiled sequence data are able to depict a comprehensive landscape of cells' stages. Furthermore, \cite{bilous_metacells_2021} has demonstrated that metacells inferred from scRNA-seq are compatible with RNA velocity model. 
Therefore, COEM is also able to be applied to mutli-omic velocity model for estimating epigenomic and transcriptomic dynamics.    

In future work, we would explore more machine learning techniques for multi-modal sequence data integration and metacell identification. For example, scATAC-seq and scRNA-seq are naturally causally related. It is interesting to utilize this unique relation to learn more robust, causally sufficient, and efficient representations~\cite{blei2020representation,wang2021desiderata,NEURIPS2021_8710ef76, Dou_Luo_Yang_2022, tabib2021myofibroblast}. Besides, the results are promising for use of an optimal transport approach to model cellular stage changes across different modalities  on the metacell level \cite{schiebinger2019optimal}.
% Similar to \citep{Persad2022.04.02.486748}, we plan to use optimal transport for cross-modality metacell integration between scATAC-seq and scRNA-seq.
\bibliography{example_paper}
\bibliographystyle{icml2022}
%%%%%%%%%%%%%%%%%%%%%%%%%%%%%%%%%%%%%%%%%%%%%%%%%%%%%%%%%%%%%%%%%%%%%%%%%%%%%%%
%%%%%%%%%%%%%%%%%%%%%%%%%%%%%%%%%%%%%%%%%%%%%%%%%%%%%%%%%%%%%%%%%%%%%%%%%%%%%%%
% APPENDIX
%%%%%%%%%%%%%%%%%%%%%%%%%%%%%%%%%%%%%%%%%%%%%%%%%%%%%%%%%%%%%%%%%%%%%%%%%%%%%%%
%%%%%%%%%%%%%%%%%%%%%%%%%%%%%%%%%%%%%%%%%%%%%%%%%%%%%%%%%%%%%%%%%%%%%%%%%%%%%%%
\newpage
\appendix
\onecolumn
\section{Appendix}

In this section, we show the results of additional experiments on datasets including Snare Cellline, 10x\_PBMC, and 10X Multiome CD34+ bone marrow.
% \vspace{-0.3in}
\begin{table}[h]
\caption{\small
Metacacell Evaluation on SNARE-seq celline dataset}
\vskip 0.1in
\begin{center}
\begin{sc}
\begin{tabular}{lcccr}
\toprule
&\textbf{RNA\_pca}&\\
\midrule
methods & separation & compactness\\
\midrule
COEM &\textbf{0.280 $\begin{tiny}{\pm 0.015}\end{tiny}$}&0.856 $\begin{tiny}{\pm 0.185}\end{tiny}$\\
SeaC-ATAC    &0.191 $\begin{tiny}{\pm 0.013}\end{tiny}$ &0.617 $\begin{tiny}{\pm 0.112}\end{tiny}$\\
SeaC-RNA    & 0.067 $\begin{tiny}{\pm 0.011}\end{tiny}$ &\textbf{0.067 $\begin{tiny}{\pm 0.004}\end{tiny}$}\\
\toprule
&\textbf{ATAC\_svd}&\\
\midrule
methods & separation & compactness\\
% COEM+linear&0.127&0.116\\
% COEM+rbf &0.121 &0.116\\
COEM &\textbf{0.159 $\begin{tiny}{\pm 0.021}\end{tiny}$}&0.124 $\begin{tiny}{\pm 0.008}\end{tiny}$\\
SeaC-ATAC    &0.097 $\begin{tiny} {\pm0.010}\end{tiny}$ &\textbf{0.004 $\begin{tiny}{\pm 0.001}\end{tiny}$ }\\
SeaC-RNA    & 0.095 $\begin{tiny}{\pm 0.018}\end{tiny}$ &0.037 $\begin{tiny}{\pm 0.012 }\end{tiny}$\\
\bottomrule
\end{tabular}
\end{sc}
\end{center}
% \vspace{-0.3in}
\label{tab:SNARE}
\end{table}
% \end{small}
\begin{small}
\vspace{-0.2in}
\begin{table}[h]
\caption{\small Metacacell Evaluation on CD34+ bone marrow dataset}
\vskip 0.1in
\begin{center}
\begin{sc}
\begin{tabular}{lcccr}
\toprule
&\textbf{RNA\_pca}&\\
\midrule
methods & separation & compactness\\
\midrule
COEM &\textbf{0.266 $\begin{tiny}{\pm 0.011}\end{tiny}$ }&0.035 $\begin{tiny}{\pm 0.006}\end{tiny}$ \\
SeaC-ATAC&0.203 $\begin{tiny}{\pm 0.013}\end{tiny}$ &0.037 $\begin{tiny}{\pm 0.007}\end{tiny}$\\
SeaC-RNA& 0.256 $\begin{tiny}{\pm 0.037}\end{tiny}$ &\textbf{0.026 $\begin{tiny}{\pm 0.003}\end{tiny}$}\\
\toprule
&\textbf{ATAC\_svd}&\\
\midrule
methods & separation & compactness\\
COEM &\textbf{0.318 $\begin{tiny}{\pm 0.014}\end{tiny}$}&0.136 $\begin{tiny}{\pm 0.010}\end{tiny}$ \\
SeaC-ATAC   &0.284 $\begin{tiny}{\pm 0.015}\end{tiny}$ &\textbf{0.025 $\begin{tiny}{\pm 0.003}\end{tiny}$ } \\
SeaC-RNA    &0.244 $\begin{tiny}{\pm 0.123}\end{tiny}$ &0.115 $\begin{tiny}{\pm 0.021}\end{tiny}$ \\
\bottomrule
\end{tabular}
\end{sc}
\end{center}
\vspace{-0.5in}
\label{tab:CD34}
\end{table}
\vspace{0.3in}
\begin{table}[ht!]
\caption{\small Metacacell Evaluation on PBMC dataset}
\vskip 0.1in
\begin{center}
\begin{sc}
\begin{tabular}{lcccr}
\toprule
&\textbf{RNA\_pca}&\\
methods & separation & compactness\\
\midrule
COEM & \textbf{0.515 $\begin{tiny}{\pm 0.127}\end{tiny}$ }&0.349 $\begin{tiny}{\pm 0.115}\end{tiny}$ \\
SeaC-ATAC&0.406 $\begin{tiny}{\pm 0.129}\end{tiny}$ &3.315 $\begin{tiny}{\pm 1.21}\end{tiny}$ \\
SeaC-RNA&0.428 $\begin{tiny}{\pm 0.133}\end{tiny}$ &\textbf{0.294 $\begin{tiny}{\pm 0.046}\end{tiny}$}\\
\toprule
&\textbf{ATAC\_svd}&\\
\midrule
methods & separation & compactness\\
COEM &\textbf{0.181 $\begin{tiny}{\pm 0.038}\end{tiny}$}&0.030 $\begin{tiny}{\pm 0.013}\end{tiny}$ \\
SeaC-ATAC   &0.169 $\begin{tiny}{\pm 0.024}\end{tiny}$ &\textbf{0.022$\begin{tiny}{\pm 0.008}\end{tiny}$} \\
SeaC-RNA    &0.034 $\begin{tiny}{\pm 0.009}\end{tiny}$&0.141 $\begin{tiny}{\pm 0.005}\end{tiny}$\\
\bottomrule
\end{tabular}
\end{sc}
\end{center}
\label{tab:10XPBMC}
\end{table}
\end{small}
\end{document}